\documentclass[final,tablecaptionbelow]
  {aipproc}
\layoutstyle{8x11single}
\def\gtrless{\raise2.5pt\hbox{$>$}\llap{\lower2.5pt\hbox{$<$}}}
\def\gtrapprox{\raise2.5pt\hbox{$>$}\llap{\lower2.5pt\hbox{$\approx$}}}

\newcommand{\bsq}[1]{\begin{subequations}\label{#1}}
\newcommand{\esq}{\end{subequations}}
\newcommand{\beq}[1]{\begin{equation}\label{#1}}
\newcommand{\eeq}{\end{equation}}
\newcommand{\beqa}[1]{\begin{eqnarray}\label{#1}}
\newcommand{\eeqa}{\end{eqnarray}}

\newcommand{\gd}{\dot{\gamma}}

\newcommand{\qb}{{\bf q}}
\newcommand{\kb}{{\bf k}}
\newcommand{\pb}{{\bf p}}

\renewcommand{\rho}{\varrho}
\renewcommand{\epsilon}{\varepsilon}

\begin{document}

\title{Viscoelasticity and shear flow of concentrated, non-crystallizing colloidal suspensions:\\ Comparison with Mode-Coupling Theory}


\author{Miriam Siebenb{\"u}rger}{
  address={Physikalische Chemie I, Universit{\"a}t Bayreuth, 95440 Bayreuth, Germany}}

\author{Matthias Fuchs} {
  address={Fachbereich Physik, Universit{\"a}t Konstanz, 78457 Konstanz, Germany}}

\author{Henning Winter}{
  address={Chemical Engineering Department, University of Massachusetts, Amherst
Box 33110, Amherst, MA 01003-3110, U. S. A.}}

\author{Matthias Ballauff }{
  address={Physikalische Chemie I, Universit{\"a}t Bayreuth, 95440 Bayreuth, Germany}}

\begin{abstract}
We present a comprehensive rheological study of a suspension of
thermosensitive particles dispersed in water. The volume fraction of
these particles can be adjusted by the temperature of the system in
a continuous fashion. Due to the finite polydispersity of the
particles (standard deviation: 17\%), crystallization is suppressed
and no fluid-crystal transition intervenes. Hence, the moduli $G'$
and $G''$ in the linear viscoelastic regime as well as the flow
curves (shear stress $\sigma(\dot{\gamma})$ as the function of the
shear rate $\dot{\gamma}$) could be measured in the fluid region up
to the vicinity of the glass transition. Moreover, flow curves could
be obtained over a range of shear rates of 8 orders of magnitude
while $G'$ and $G''$ could be measured spanning over 9 orders of
magnitude. Special emphasis has been laid on precise measurements
down to the smallest shear rates/frequencies. It is demonstrated
that mode-coupling theory generalized in the integration through
transients framework provides a full description of the flow curves
as well as the viscoelastic behavior of concentrated suspensions
with a single set of well-defined parameters.
\end{abstract}

\maketitle


\section{Introduction}
The flow behavior of concentrated suspensions of hard sphere particles in a shear flow is a long-standing problem in colloid physics [Russel \textit{et al.} (1989), Larson (1999)]. The most prominent feature of these systems is the strong increase of the zero-shear viscosity with increasing volume fraction $\phi$. Concomitantly, there is a strong increase of the storage modulus $G'$. Moreover, with increasing shear rate a marked shear thinning sets in where the relative viscosity $\eta/\eta_S$ ($\eta_S$: viscosity of solvent) is lowered considerably. These general features are also observed for solutions of soft colloidal particles as e.g. a solution of star polymers in a suitable solvent [Helgeson \textit{et al.} (2007)]. Suspensions of hard spheres or of particles characterized by a sufficiently steep mutual repulsion may hence be treated as the most simple model system for the rheological study of the flow behavior of colloidal systems.

By now, it is well understood that the dynamics of quiescent
suspensions of hard spheres can be understood and modeled
quantitatively in terms of the mode-coupling theory  [van Megen and
Pusey (1991), van Megen and Underwood (1994)]. Mode-coupling theory
 [G{\"o}tze and Sj{\"o}gren (1992)] was shown to explain
quantitatively the structural arrest that takes place in colloidal
dispersions above the volume fraction $\phi_g \approx 0.58$.
Moreover, this value of $\phi_g$ has turned out to be in full
agreement with experimental studies of the zero-shear viscosity
$\eta_0$ [Meeker  \textit{et al.} (1997), Cheng \textit{et al.}
(2002)]. Here it has been demonstrated that the strong upturn of
$\eta_0$ occurs much below the volume fraction of 0.63 that would
refer to random close packing of the spheres. First theoretical
calculations of $\eta_0$ within the frame of MCT fully supported
this view [Banchio \textit{et al.} (1999)].

Fuchs and Cates (2002, 2003) recently demonstrated that MCT can be
generalized to treat the dynamics of colloidal suspension in shear
flow. They developed the integrations throught transients (ITT)
approach, which describes the competition between enforced
deformation or flow and intrinsic structural relaxation.  Thus, MCT
can be used to model the distortion of the structure of suspensions
by a flow field at finite shear rate $\dot{\gamma}$. One central
point of this theory is the melting of the glass by shear: flow
leads to a glass-to-fluid transition in the suspension and a unique
stationary state can be reached again despite the fact that the
volume fractions of the particles in the suspension are above
$\phi_g$. It is presumed that this holds as long as local 'free
volume' is available below random close packing. Flow curves giving
the relative viscosity $\eta/\eta_S$ as function of shear rate are
predicted to present well-defined states that are independent of the
previous history of the sample. In addition to these predictions
from ITT-MCT pertaining to the non-linear flow of suspensions, MCT
is capable of describing the linear-viscoelastic behavior of
concentrated dispersions. In this way MCT offers a complete modeling
of the rheological  properties of dispersions at high packing
fractions $\phi$ or for strong particle interactions close to
vitrification. First comparisons with simulations indeed
demonstrated that the theory captures all of the salient features of
the flow behavior of flowing suspensions [Varnik and Henrich
(2006)].

\begin{figure}[t]
\centering
\includegraphics[width=0.7\textwidth]{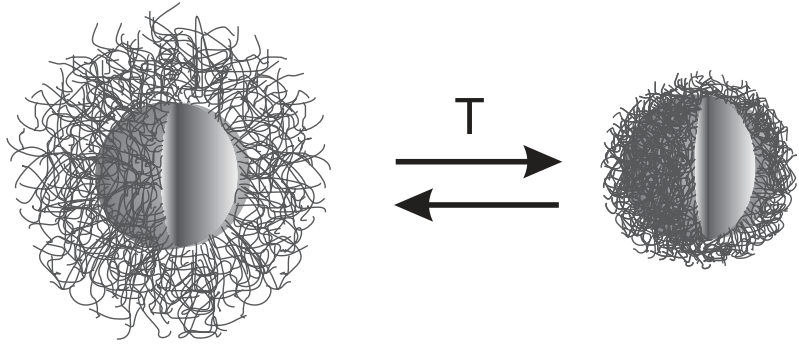}
\caption{Schematic representation of the thermosensitive core-shell particles used in this study: The polymer chains are affixed to the
surface of the core. Immersed in cold water the thermosensitive network will swell. The degree of swelling can be reduced by raising the temperature. At 32$\,^o$C a volume transition takes place in which most of the water is expelled from the network. }
\label{Fig:1}
\end{figure}

Recently, it has been demonstrated that suspensions of
thermosensitive particles present excellent model systems for
studying the flow of concentration suspensions for a quantitative
comparison of the theory of Fuchs and Cates to experimental data
[Fuchs and Ballauff (2005), Crassous \textit{et al.} (2006a, 2008a)]. The particles consist of a solid
core of polystyrene onto which a thermosensitive network of
poly(N-isopropylacrylamide) (PNIPAM) is attached ([Crassous
\textit{et al.} (2006b, 2008b)], see Figure \ref{Fig:1}). The PNIPAM
shell of these particles swells when immersed in cold water (below 32$\,^o$C). Water gets expelled at higher temperatures leading to a
considerable shrinking. Thus, for a given number density the
effective volume fraction $\phi_{eff}$ can be adjusted within wide
limits by adjusting the temperature. Senff \textit{et al.} (1999)
were the first to demonstrate the use of these particles as model
system for studying the dynamics in concentrated suspensions [Senff
\textit{et al.} (1999)]. Similar microgel particles  have been used
to study the dynamics of concentrated suspensions and the aging in
these systems [Purnomo \textit{et al.} (2006, 2007)]. The advantage of these systems over the classical hard
sphere systems are obvious: Dense suspensions can be generated
\textit{in situ} without shear and mechanical deformation. Moreover,
all previous history of the sample can be erased by raising the
temperature and thus lowering the volume fraction to the fluid
regime. Hence, subtle effects as e.g. aging can be studied by these
systems in a nearly ideal fashion [Purnomo \textit{et al.} (2006, 2007)]. Moreover, microgels can be used as
models for pastes and gels in which the effective volume fraction of
the particles exceeds unity by far [Seth \textit{et al.} (2006)].

Previous investigations using dispersions of thermosensitive
core-shell particles have demonstrated that mode-coupling theory is
indeed able to model the flow curves as well as the linear
viscoelasticity of concentrated suspensions [Crassous \textit{et
al.} (2006a, 2008a)]. In particular, this
theory allows us to calculate $G'$ and $G''$ as function of the
frequency $\omega$ from data that have been obtained from the
analysis of flow curves [Crassous \textit{et al.} (2008a)]. However,
a problem that restricted this comparison was crystallization due to
the narrow size distribution of the particles. Crystallization
presents a major obstacle since the rate of nucleation is rather
fast for volume fractions even far below the volume fraction
$\phi_g$ of glass transition ($\phi_g \approx 0.58$). The
crystalline phase could be removed by shear melting and the fluid
state could be super-cooled for a given time. The time window of
nucleation, however, limited the experimental studies to frequencies
and/or shear rates not too small [Crassous \textit{et al.} (2006a, 2008a)]. A fully conclusive comparison of
MCT with experimental data, however, requires that measurements
extend down to the smallest frequencies and shear rates [Fuchs and
Ballauff (2005), Crassous \textit{et al.} (2008a)].

Here we present a full comparison of theory and experiment using a thermosensitive suspension having a slightly broader size distribution that prevents crystallization. Hence, all measurements can now be extended to very low shear rates or frequencies that may take hours without any disturbance by crystallization. It will be demonstrated that MCT provides indeed a full description of the rheological behavior of suspension under homogeneous flow. Special attention will be paid to shear banding  [Callaghan (2008), Dhont and Briels (2008), Manneville (2008), Olmsted (2008)] and other effects that may come into play for concentrated suspensions [Isa \textit{et al.} (2007)] and disturb the quantitative comparison of theory and experiment.

The paper is organized as follows: In the following section we give
a brief survey of the main predictions of MCT. In particular, the
equations used for the evaluation of data as well as the fit
parameters necessary for a comparison of MCT with experimental data
are discussed. For a extensive description of MCT the reader is
deferred to a recent paper [Crassous \textit{et al.} (2008a)].
After a section devoted to the description of the systems and the
measurements, a comprehensive comparison of MCT with flow curves as
well as with linear viscoelastic properties will given. A major
focus is the low frequency and long time behavior which had not
been accessible before. Special attention will be paid to the fact
that MCT is capable of describing both the flow curves as well as
$G'$ and $G''$ as functions of frequency. A brief Conclusion will
sum up the main findings.

\section{Mode-coupling theory of flowing suspensions: Main predictions}

The central points and main predictions of MCT and its
generalization to flowing dispersions in the integration through
transients (ITT) approach have recently been discussed in detail
[Crassous \textit{et al.} (2008a)]. Hence, it will suffice to
discuss the points that will be used in the analysis of the
experiments. Structural relaxation, which is the central topic of
MCT, leads to internal relaxation processes with an intrinsic
relaxation time $\tau$ that far exceeds the dilute limit Brownian
diffusion time. The ITT-MCT approach to sheared fluids considers the
effect of shear when the dressed Peclet or Weissenberg number $Pe =
\dot\gamma \tau$ becomes appreciable, while  the bare Peclet number
$Pe_0 = 6\pi \eta_s R_H^3 \dot\gamma/(k_BT)$ formed with the solvent
viscosity $\eta_s$ and the hydrodynamic particle radius $R_H$ should
be much lower than unity for the approach to hold.

The main point of the MCT of quiescent dispersions is the slowing
down of the dynamics at the glass transition.  Critical values for
the particle concentration, the temperature or other thermodynamic
control variables exist, where a glass transition separates  fluid
from yielding glassy states. In fluid states, a first Newtonian
region is reached if the shear rate or the frequency is low enough;
the Newtonian viscosity $\eta_0=\eta(\dot\gamma\to0)$ is finite.
Non-Newtonian flow, that is shear thinning, is observed at higher
shear rates corresponding to a finite Weissenberg number Pe $\approx
1$. MCT recovers Maxwell's relation $\eta_0\propto \tau$, so that
the criterion $Pe \approx 1$ translates into
$\eta_0\dot\gamma\approx G^c_\infty$, where
$G^c_\infty=G'(\omega\to0)$ is the (zero-frequency) shear modulus at
the glass transition. For glassy states at higher densities or
interaction parameters, Newtonian flow is not observed anymore but
the shear stress $\sigma$ is predicted to approach a plateau value
for sufficiently low shear rates $\dot{\gamma}$; a dynamic yield
stress follows $\sigma^+=\sigma(\dot\gamma\to0)>0$. It corresponds
to an infinite Newtonian viscosity.

A central point of MCT as applied to flowing suspensions [Fuchs and
Cates (2002)] is shear melting: Finite shear rates will lead to a
melting of the glass into a fluid. As a consequence of this, flow
curves given by the shear stress $\sigma$ as function of the shear
rate $\dot{\gamma}$ present well-defined states for a given volume
fraction even above the glass transition. Hence,
$\sigma(\dot{\gamma})$ is independent of the previous history of the
sample, under the condition that the stationary long time limit has
been achieved. Previous experimental studies strongly suggest that
this central prediction of theory is found indeed [Crassous
\textit{et al.} (2006a, 2008a)]. However,
in all previous experiments crystallization intervened at
sufficiently low shear rates and this point is in need of further
experimental elucidation.

In the following, the main equations used for the evaluation of the
rheology data within MCT are briefly summed up. Within the
\textit{schematic $F_{12}^{(\dot{\gamma})}$-model} [Fuchs and Cates
(2003)] Brownian motion of the particles is captured in terms of a
density correlator $\Phi(t)$ which is normalized/initiated
according to $\Phi(t\to0)= 1-\Gamma t$. MCT relates this correlation
to the memory function $m(t)$ by a memory equation
\begin{equation}
\label{d1}
\partial_t \Phi(t) + \Gamma \left\{ \Phi(t) + \int_0^tdt'\; m(t-t') \;
\partial_{t'} \Phi(t') \right\}  = 0 \; .
\end{equation}
The parameter $\Gamma$ describes the microscopic dynamics of the density correlator and will depend on
structural and {\em hydrodynamic} correlations. The memory function
describes stress fluctuations which become slower when approaching the volume fraction of the glass transition. In the
$F_{12}^{(\dot{\gamma})}$-model the memory function is written as
\begin{equation}
\label{d2}
m(t)= \frac{v_1 \, \Phi(t) + v_2\, \Phi^2(t)}{1+\left(\dot\gamma
t/\gamma_c\right)^2}.
\end{equation}
This model, for the quiescent case $\dot\gamma=0$, had been
suggested by G\"otze in 1984, [G\"otze (1984, 1991)] and
describes the development of slow structural relaxation upon
approaching a glass transition; it is reached by increasing  the
coupling vertices/ parameters $v_1$ and $v_2$ to their critical
values $v_1^c=(2\lambda-1)/\lambda^2$ and $v_2^c=1/\lambda^2$, where
the so-called exponent parameter $\lambda$ with $1/2\le \lambda \le
1$ parametrizes the (type-B) glass transition line. Under shear an
explicit time dependence of the couplings in $m(t)$ captures the
accelerated loss of memory by shear advection [Fuchs and Cates
(2002, 2003, 2008)]. Shearing causes the dynamics to decay for long
times, because fluctuations are advected to smaller wavelengths
where small scale Brownian motion relaxes them. The parameter
$\gamma_c$ sets the scale for the accumulated strain $\dot\gamma t$
to matter.

 Equations (\ref{d1}, \ref{d2})
will be used with  the choice of vertices $v_2=v_2^c=2$ (corresponding to $\lambda=1/\sqrt2$), and
$v_1=v_1^c+\epsilon\,(1-f^c)/f^c$, where $v_1^c=0.6258$. The critical glass form factor $f^c$,  which takes the value $f^c=(1-\lambda)= 0.293$ for this choice, describes the frozen-in structure obtained from $\Phi(t\to\infty,\varepsilon=0,\dot\gamma=0)=f^c$ at the glass transition in the absence of shear. The glass transition lies at $\varepsilon=0$, where the Newtonian viscosity diverges. Under shear, glass states are molten, but the transition still separates shear thinning fluid states, $\varepsilon<0$, where a Newtonian viscosity exists, from yielding glassy states $\varepsilon\ge0$, where a dynamic yield stress $\sigma^+$ exists. The specific choice of this transition point is motivated by $(i)$ its repeated use in the literature, and $(ii)$ because it predicts a divergence of the viscosity at the glass transition, whose power is close to the one of the microscopic MCT calculation for short ranged repulsions:
$$\eta_0 \propto \tau \propto \left( - \varepsilon \right)^{-\gamma}\; ,\quad \mbox{with }\; \gamma=\gamma(\lambda=1/\sqrt2) = 2.34\; .$$
Within microscopic MCT for hard spheres using the Percus-Yevick structure factor $\gamma=2.46$ was found [G\"otze and Sj\"ogren (1992)].

The correlator $\Phi(t)$ obtained by numerical solution of Eq. \ref{d1} may then be used to calculate the generalized shear modulus $g(t,\dot\gamma)$ according to
\begin{equation}
\label{d4}
g(t,\dot\gamma)= v_\sigma \,\Phi^2(t) + \eta_\infty \; \delta(t-0+) \; .
\end{equation}
The parameter $\eta_\infty$ characterizes a short-time high
frequency viscosity and models viscous processes which require no
structural relaxation. Together with $\Gamma$,
it is the only model parameter affected by solvent mediated
interactions. Steady state shear stress under constant shearing, and
viscosity then follow via integrating up the generalized modulus:
\begin{equation}
\label{d5}
 \sigma = \eta \; \dot\gamma =  \dot{\gamma}\; \int_0^\infty\!\!\!
dt\; g(t,\dot\gamma)  = \dot{\gamma}\; \int_0^\infty\!\!\!dt\;v_\sigma
\Phi^2(t) + \dot{\gamma}\; \eta_\infty \; .
\end{equation}
When setting the shear rate $\dot\gamma=0$ in Eq.~\ref{d2} so that the schematic correlator belongs to the quiescent equilibrium system, the frequency dependent moduli of the linear response framework follow from Fourier transforming:
\begin{equation}
\label{d6}
G'(\omega) + i\, G''(\omega)  = i\, \omega \, \int_0^\infty\!\!\!
dt \; e^{-i\, \omega\, t}\;g(t,\dot\gamma=0)  =  i\, \omega \,
\int_0^\infty dt\; e^{-i\, \omega\, t}\;  v_\sigma \left.
\Phi^2(t)\right|_{\dot\gamma=0} + i \omega\, \eta_\infty\;
\end{equation}
Because of the vanishing of the Fourier-integral in Eq.~\ref{d6} for high
frequencies, the parameter $\eta_\infty$ can be directly identified as high
frequency viscosity:
\begin{equation}
\label{d7}
\lim_{\omega\to\infty} G''(\omega) / \omega
=\eta^{\omega}_\infty\quad , \; \mbox{with }\;\eta^{\omega}_\infty =
\eta_{\infty}\; .
\end{equation}
At high shear, on the other hand, Eq.~\ref{d2} leads to a vanishing of
$m(t)$, and Eq.~\ref{d1} gives an exponential decay of the transient
correlator, $\Phi(t) \to e^{-\Gamma\, t}$ for $\dot\gamma\to0$. The high
shear viscosity thus becomes
\begin{equation}
\label{d8}
\eta^{\dot{\gamma}}_\infty =
\lim_{\dot{\gamma}\to\infty} \sigma(\dot{\gamma})/ \dot\gamma = \eta_\infty +
\frac{v_\sigma}{2\, \Gamma} = \eta^\omega_\infty +
\frac{v_\sigma}{2\, \Gamma} \; .
\end{equation}
Eq. \ref{d8}  connects the high-shear viscosity with the high-frequency limit of the viscosity in an analytical manner. Moreover, both $\eta^{\dot{\gamma}}_\infty$ and $\eta^\omega_\infty$ are accessible from experimental data and Eq. \ref{d8} provides an additional check for internal consistency of the fit within the frame of the schematic model. Representative solutions of the $F_{12}^{(\dot\gamma)}$-model have been discussed recently [Crassous et al. (2008a), Fuchs and Cates (2003), Hajnal and Fuchs (2008)].

\section{Experimental}
The particles consist of a solid core of poly(styrene) onto which a network of crosslinked poly(N-isopropylacrylamide) (PNIPAM) is affixed [Crassous \textit{et al.} (2006b, 2008b)]. The degree of crosslinking of the PNIPAM shell effected by the crosslinker N,N'-methylenebisacrylamide (BIS) was 2.5 Mol $\%$. The core-shell type PS-PNIPAM particles were synthesized, purified and characterized as described in [Dingenouts \textit{et al.} (1998)]. Immersed in water the shell of these particles is swollen at low temperatures. Raising the temperature above 32$\,^o$C leads to a volume transition within the shell. Cryogenic transmission electron microscopy and dynamic light scattering have been used to investigate the structure and swelling of the particles [Crassous \textit{et al.} (2006b, 2008b)].
Further experimental details on cryogenic transmission electron microscopy are given in [Crassous \textit{et al.} (2006b)].\\
The average size and size distribution of the particles was determined with the CPS Disc Centrifuge from LOT, model DC 24000 in a aqueous glucose density gradient between 3 and 7$\,wt$\% as medium at room temperature. The number-average diameter determined at 25$\,^oC$ is 131$\,nm$ and the standard deviation of the size distribution is 17$\,$\%. \\
Dynamic light scattering (DLS) was done with a Peters ALV 800 light scattering goniometer. The temperature was controlled with an accuracy of 0.1 $^{o}$C. In the range of temperatures used for the measurements, $R_H$ could be linearly extrapolated between 6 and 25$^{o}$C ($R_H$ = -0.7796 $\cdot$T +102.4096 with T the temperature in $^{o}$C) (see [Crassous \textit{et al.} (2006a)]). Fig. \ref{Fig:2}b displays $R_H$ as the function of temperature. Moreover, this plot demonstrates that $R_H$ can obtained very precisely. Since the subsequent reduction of the data (see below) involves a scaling of the data by $R_H^3$, this fact is of great importance.

Density measurements have been made using a DMA-60 densiometer supplied by Paar (Graz, Austria). The density for the core-shell particles were determined as follows: $\rho$ = 1.1184$\,g\,cm^{-3}$ at 25$\,^{o}C$, 1.1239$\,g\,cm^{-3}$ at 20$\,^{o}C$, 1.1246$\,g\,cm^{-3}$, at 18$\,^{o}C$ and 1.1259$\,g\,cm^{-3}$ at 17$\,^{o}C$. These densities can be described with the function $\rho = 6.800 \cdot 10^{-7} \cdot T^3 + 1.129$ with the temperature T in $^{o}C$.
In all experiments reported herein 5.10$^{-2}\,mol\,L^{-1}$ KCl were added to the suspensions in order to screen the remaining electrostatic interactions caused by a few charges attached to the surface of the core particles. Previous studies have shown that this screening is fully sufficient. Thus, within the range of volume fractions under consideration here the thermosensitive particles can be regarded as hard spheres (see the discussion of this point in [Crassous \textit{et al.} (2006a)]).

The flow behavior and the linear viscoelastic properties for the range of the low frequencies were measured with a stress-controlled rotational
rheometer MCR 301 (Anton Paar), equipped with a Couette system (cup diameter: 28.929$\,mm$, bob diameter: 26.673$\,mm$, bob length: 39.997$\,mm$). For comparison the double gap geometry (inner cup diameter: 23.820$\,mm$, inner bob diameter: 24.660$\,mm$, outer bob diameter: 26.668$\,mm$, outer cup diameter: 27.590$\,mm$ and bob length: 40.00$\,mm$) was used as well.
Measurements have been performed on 12$\,mL$ (Couette single gap geometry) and 4$\,mL$ (Couette double gap geometry) solution and the temperature was set with an accuracy of $\pm$ 0.05$\,^o$C.
Deformation tests were done from a deformation of 0.1 to 100$\,$\% in the case of the dynamic deformation test with a logharitmic time ramp of 200 to 20$\,s$ at 1Hz.
The shear stress $\sigma$ versus the shear rate $\dot{\gamma}$ (flow curve) was measured after a pre-shearing of $\dot{\gamma} = 100\,s^{-1}$ for two minutes and a waiting time of 10$\,s$. The flow curves were obtained setting $\dot{\gamma}$ to a given value, first with increasing $\dot{\gamma}$ from $\dot{\gamma} =10^{-5}-10^3\,s^{-1}$ with a logarithmic time ramp from 1500 to 20$\,s$. Then after a waiting time of 10$\,s$ $\dot{\gamma}$ was decreased again.

The frequency dependence of the loss $G''$ and elastic $G'$ moduli has been measured after a two minute pre-shear at a shear rate of 100$\,s^{-1}$ and 10$\,s$ waiting time for 1$\,\%$ strain from 10 to 10$^{-4}\,Hz$ with a logarithmic time ramp from 20 to 1500$\,s$. The dependence upon the strain has been checked and confirmed that all the measurements were performed in the linear regime.
$G'$ and $G''$ at elevated frequencies were obtained using the Piezoelectric vibrator (PAV) [Crassous \textit{et al.} (2005)] and cylindrical torsional resonator [Deike \textit{et al.} (2001), Fritz \textit{et al.} (2003)] supplied by the Institut f{\"u}r dynamische Materialpr{\"u}fung, Ulm, Germany.  The PAV was operated from 10 to 3000$\,Hz$. The solution is placed between two thick stainless steel plates. One remains static whereas the second is cemented to piezoelectric elements. The gap was adjusted with a 50$\,\mu\,m$ ring. A set of piezoelectric elements is driven by an ac
voltage to induce the squeezing of the material between the two plates. The second set gives the output voltage. Experimental details concerning this instrument are given elsewhere [Crassous \textit{et al.} (2005)].
The cylindrical torsional resonator (TR) used was operated at a single frequency (26$\,kHz$). The experimental procedure and the evaluation of
data have been described recently [Deike \textit{et al.} (2001), Fritz \textit{et al.} (2003)].  \\

\section{Results and Discussion}

\subsection{Characterization}

\begin{figure}[t]
\centering
\includegraphics[width=0.95\textwidth]{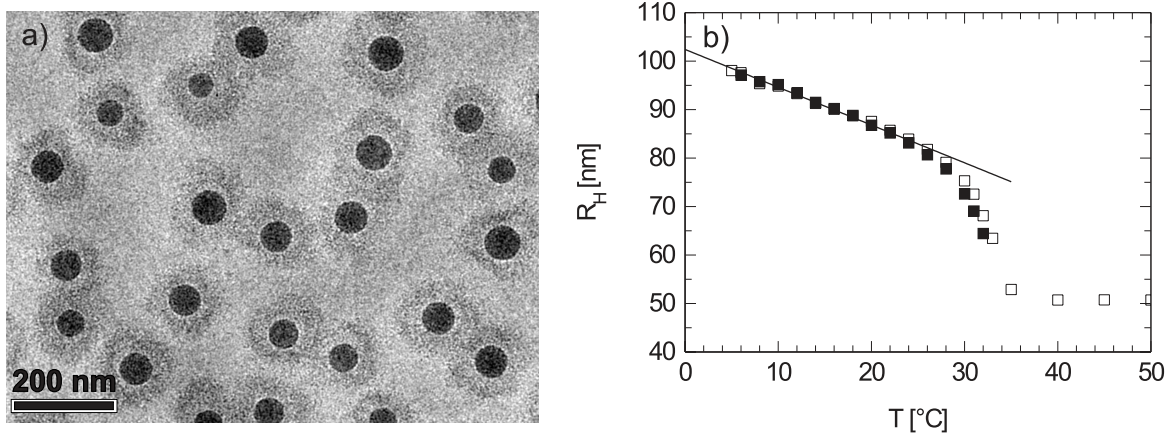}
\caption{a) Cryogenic Transmission Electron Microscopy micrograph of the core-shell particles in water. b) Dynamic light scattering measurements of the latex at varius temperatures. Filled squares represent the hydrodynamic radius measured in aqueous 0.05$\,mol\,L^{-1}$ KCl, hollow squares measured in water. All rheological experiments presented here have been done using particles dissolved in aqueous 0.05$\,mol\,L^{-1}$ KCl solution [Crassous \textit{et al.} (2006a, 2008a)]. In this way possible electrostatic interaction between the particles due to the residual charges affixed to the core particles can be screened totally. Below the volume transition the decrease of the hydrodynamic radius $R_H$ giving the outer boundary of the network is linear in very good approximation ($R_H$ = -0.7796 $\cdot$T +102.4096 with T being the temperature in $^{o}$C; see solid line in graph). \label{Fig:2} }
\end{figure}

Recent work has shown that Cryo-TEM is a excellent tool for the analysis of the core-shell particles used for the present study [Crassous \textit{et al.} (2006b, 2008b)] because this method allows us to visualize the particles in-situ, that is, directly in the aqueous suspension.  Fig. \ref{Fig:2} displays a micrograph of the suspension obtained by cryo-TEM. The core-shell structure of the particles is clearly visible as already demonstrated by previous work [Crassous \textit{et al.} (2006b, 2008b)]. Since the polydispersity of the present suspension is considerably higher than the one of the dispersion used in our previous study [Crassous \textit{et al.} (2008a)], the system does not crystallize anymore even if kept for a prolonged time at volume fractions above 50$\,\%$. However, Fig. \ref{Fig:2} demonstrates that the suspension used here is still a well-defined model system. Hence, it can be used for a comprehensive comparison of experimental data with MCT. Attempts to suppress crystallization by mixing core-shell particles of slightly different size failed to produce reliable results since crystallization in these systems was found to be only slowed down but not suppressed totally.

The effective volume fraction $\phi_{eff}$ could not be calculated
as for monodisperse particles by precise measurements of the
hydrodynamic radius $R_H$, the core radius and the mass ratio of
core and shell as done in [Crassous \textit{et al}. (2008a)].
Unfortunately, polydispersity leads to an error which is too large
and no meaningful data could be obtained. Furthermore, viscosity
measurements in the dilute state and the use of the
Batchelor-Einstein equation for the relative zero shear viscosity could not be applied because of the
appreciable polydispersity, too. Therefore the effective volume
fraction was obtained by comparing the relative viscosities at
infinite shear rate with the monodisperse core-shell system used in
[Crassous \textit{et al}. (2008a)]. This will be discussed in further
detail in conjunction with Fig. \ref{Fig:6} below.

\subsection{Reversibility of flow curves}

\begin{figure*}
    \centering
    \includegraphics[width=1\textwidth]{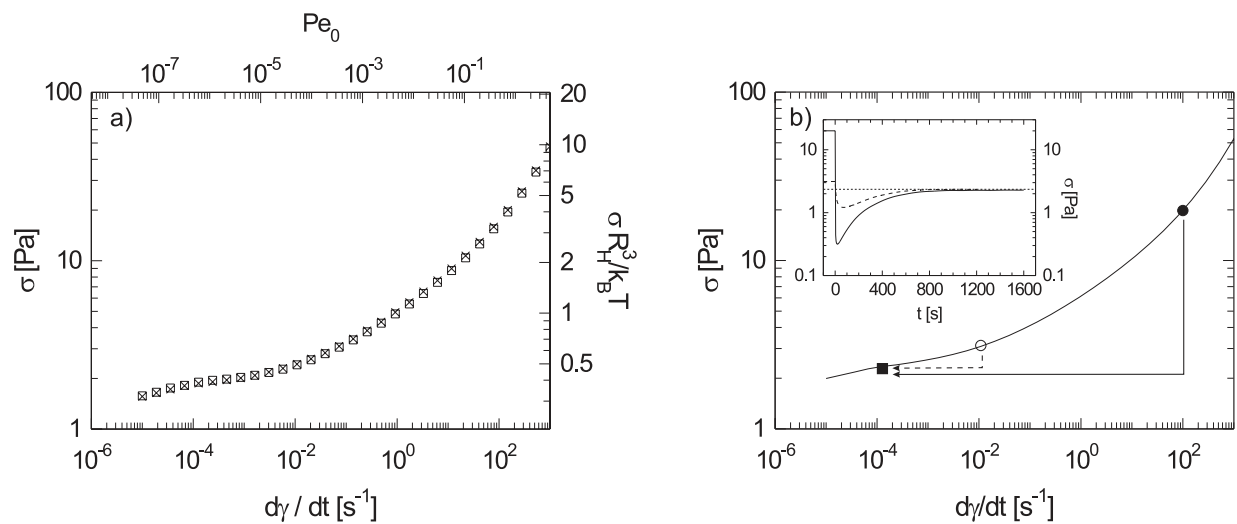}
    \caption{a) Reversibility of the flowcurves for a 7.87$\,wt\%$ suspension at 12$\,^{\circ}C$ with two different geometries: Couette single gap - geometry (crosses) and double gap - geometry (squares). b) Reversibilty of the flowcurves by comparing jump experiments and the flowcurve (solid line) measured from higher to lower shear rates at 8.42$\,wt\%$ and 14.75$^{\circ}C$. The steady state shear stress values of jumps from different shear rates to one as shown by the arrows in the graph are plotted for two examples: jump from 100$\,s^{-1}$ (full circle) to 0.000127$\,s^{-1}$ (full square) and jump from 0.0108$\,s^{-1}$ (hollow circle) to 0.000127$\,s^{-1}$ (full square). b) Inset: the same jumps are plotted time-dependent. The different jumps are plotted like followed: jump from 100$\,s^{-1}$ to 0.000127$\,s^{-1}$ (thick solid line) and jump from 0.0108$\,s^{-1}$ to 0.000127$\,s^{-1}$ (thick dashed line). The thin short dashed line in the inset represents the shear stress value at 0.000127$\,s^{-1}$ of the flowcurve for the same concentration and temperature.
        \label{fig:3} }
\end{figure*}

A central prediction of MCT is the full reversibility of the flow curves, that is, of the shear stress $\sigma$ as the function of the shear rate $\dot{\gamma}$. This is evident for any volume fraction below the volume fraction $\phi_g$ of the glass transition. Here the system is ergodic and a well-defined steady state must be reached for sufficiently long time. This point is well borne out of all experiments done so far [Crassous \textit{et al.} (2006a, 2008a)]. However, a totally different situation arises above $\phi_g$. In the glassy state, the suspension becomes non-ergodic and the flow curves may become ill-defined states. Hence, for concentrated suspensions of large non-Brownian particles dispersed in a highly viscous medium, the measured shear stress is found to be dependent on the previous history of the sample [Heymann and Aksel (2007)]. MCT, however, predicts that shear melting sets in at any finite shear rate $\dot{\gamma}$. Therefore, shear flow restores the ergodicity in the system and thermal motion of the spheres can erase all traces of the previous deformations of the sample if flow is applied over a sufficiently long time interval. Two types of experiments need to be done to explore these basic principles in further detail:\\

i) Each data point giving $\sigma(\dot{\gamma})$ must be independent
regardless of the shear history. Thus, flow curves
$\sigma(\dot{\gamma})$ should be fully reversible regardless whether
the data are taken by increasing rate or decreasing the shear rate.
The reversibility of the flow curves has already been demonstrated
in previous work. However, the suspensions analyzed previously
[Crassous \textit{et al.} (2006a, 2008a)] began to crystallize during the
long measurements needed for exploring the dynamics at smallest
shear rates $\dot{\gamma}$. To overcome this problem, the present
system is sufficiently polydisperse, crystallization is suppressed.
Fig. \ref{fig:3} displays the resulting flow curves down to $Pe_0 =
10^{-6}$. The onset of a plateau is clearly seen. Additional
experiments with long waiting times demonstrated that
crystallization does not occur in the present system. Moreover, Fig.
\ref{fig:3}a points to another important fact:
$\sigma(\dot{\gamma})$ is entirely independent of the geometry of
the rheometer used for studying the flow behavior. Fig. \ref{fig:3}a
demonstrates this fact by overlaying the resulting
$\sigma(\dot{\gamma})$ obtained from a single gap and double gap
Couette geometry. If shear banding would play any role [Callaghan
(2008), Dhont and Briels (2008), Manneville (2008), Olmsted (2008)],
the resulting $\sigma(\dot{\gamma})$ should be different and depend
definitely on the width of the gap of the instrument. The difference
in curvature for the two experiment geometries would have resulted
in different onsets of shear banding. Also, shear banding in the
double-gap Couette system would have resulted in a complex interplay
of instabilities in the two gaps. None of this is observed here.

ii) The full reversibility of the flow curves are demonstrated also by a second type of experiment (Fig. \ref{fig:3}b): Here a stationary state is first obtained for a given shear rate $\dot{\gamma}$. Then a much lower shear rate is adjusted and the resulting shear stress  is monitored as a function of time. Fig. \ref{fig:3}b shows that $\sigma(\dot{\gamma})$  converges to the stationary value measured previously (solid line in Fig. \ref{fig:3}b). The inset displays the evolution of $\sigma(\dot{\gamma})$ with time. Evidently, a long time may be needed until the stationary value is reached. However, all data demonstrate that the full reversibility of $\sigma(\dot{\gamma})$ is fully supported by our experiment also above the glass transition. One of the central predictions is thus fully corroborated.  Deviations observed previously [Crassous \textit{et al.} (2008a)] were solely due to crystallization of the sample which is suppressed in the present system. In this way flow curves may to a certain extent be compared to thermodynamic results, they are fully independent of the path leading to a given state.

\begin{figure*}
    \centering
        \includegraphics[width=0.5\textwidth]{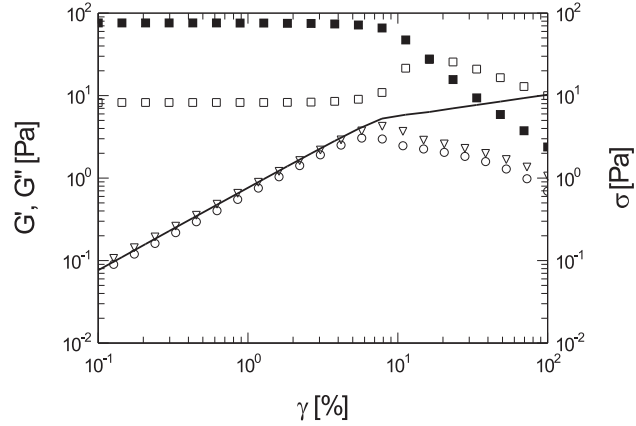}
    \caption{Comparison of the dynamic strain sweep at 1$\,Hz$ (\textit{G'}: filled squares, \textit{G''} open squares, $\sigma$ solid line) and a continuous strain sweep with two different measurement times for each point (20$\,s$ per point: triangles down and 200$\,s$ per point: circles) for a 8.90$\,wt\%$ suspension at 17$\,^{\circ}$C ($\phi_{eff}$ = 0.627).
    \label{fig:4}}
\end{figure*}

A point to be considered further is the transition from the linear
to the non-linear viscoelastic regime in the glassy regime. Fig.
\ref{fig:4} displays $G'$, $G''$ as well as $\sigma$ for different
deformations at a rate of 1 Hz. The data show that the linear regime
is preserved up to a strain of nearly 10$\%$. Moreover, $G'\gamma$
measured for different strains $\gamma$ coincides directly with the
measured stress $\sigma$. Hence, within this range of $\gamma$, the
glassy suspensions behaves as a homogeneous solid as expected.
Again, shear banding or possible inhomogeneities within the gap
should disturb this type of measurement. Fig. \ref{fig:4}
demonstrates that this is not seen. Obviously, these experiments are
not fully sufficient to rule out shear banding entirely. However,
the good agreement between two entirely different types of
experiments with theory to be discussed further below points clearly
to the fact that shear banding does not disturb the measurements
reported here.

Beyond a strain of ca. 10 $\%$, shear melting of the glass starts
and the measured stress $\sigma$ is considerably lower than the
value calculated from $G'$. Fig. \ref{fig:4}  indicates that the
linear viscoelastic regime is extended over a sufficiently large
range of deformations.

\subsection{Comparison with mode coupling theory: Schematic model}
A thorough comparison of ITT-MCT to experimental data must be related not only to the flow curves discussed in the previous section but also to data obtained for the linear viscoelasticity, that is for $G'$ and $G''$ as  function of the frequency $\omega$. A previous comparison [Crassous \textit{et al.} (2008a)] has demonstrated that both sets of data can indeed be described by a generalized modulus $g(t,\dot\gamma)$ as introduced in the section Theory. However, at smallest Pe$_0$ crystallization intervened and the comparison was restricted to intermediate to high shear rates/frequencies. Crystallization is suppressed for the present system and a fully quantitative comparison of theory and experiment now becomes possible down to $Pe_0 \approx 10^{-8}$ in case of the flow curves. $G'$ and $G''$ can be obtained down to  $Pe_{\omega} \approx 10^{-6}$. Here, the frequency dependent Peclet (or Deborrah) number, defined as $Pe_\omega  = 6\pi \eta_s R_H^3 \omega/(k_BT)$, compares the frequency with the time it takes a particle to diffuse its radius $R_H$ at infinite dilution. At high densities, structural relaxation shows up at Pe$_\omega \ll 1$ .

\begin{figure*}
    \centering
    \includegraphics[width=0.95\textwidth]{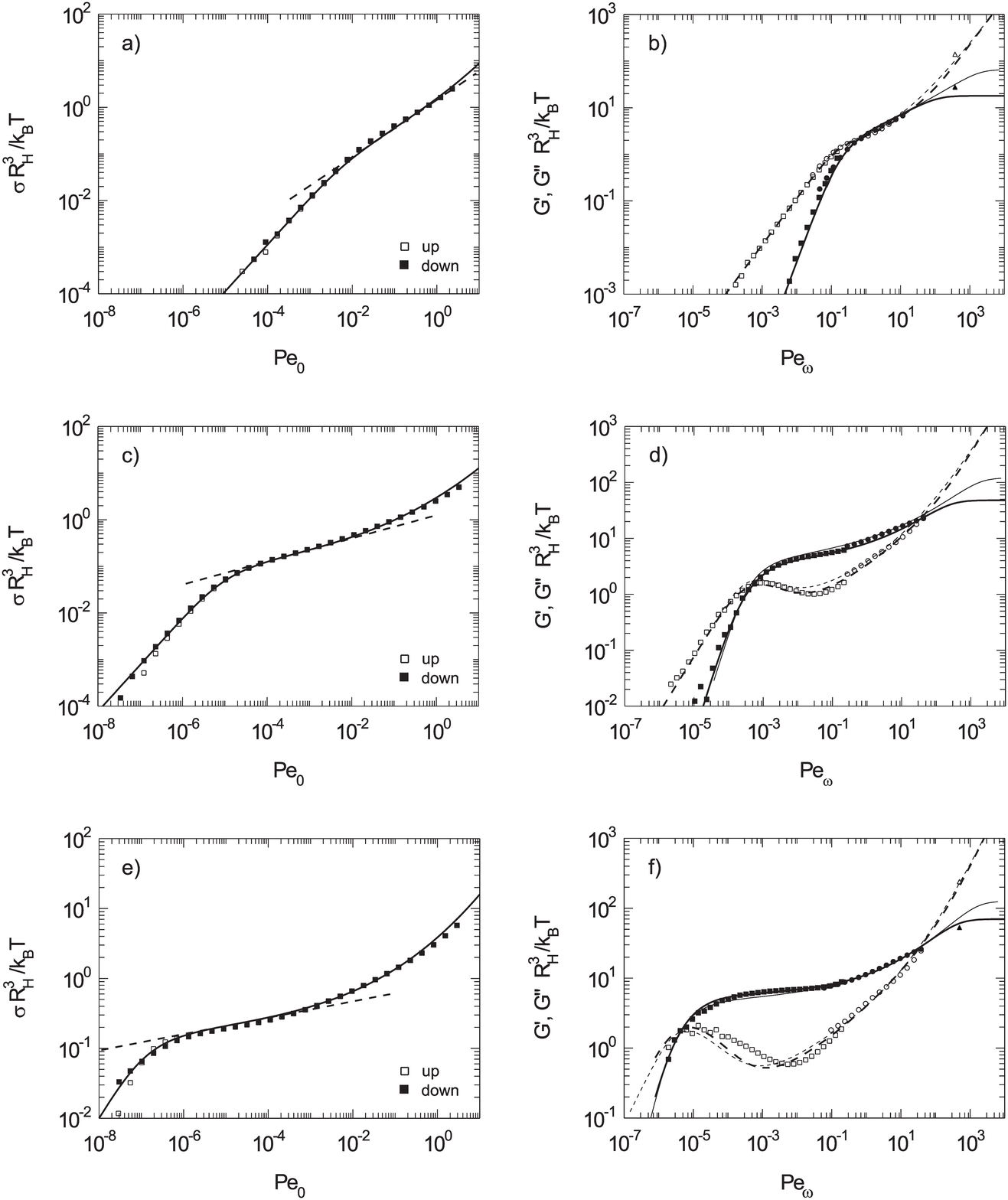}
\caption{Left column: Reduced flow curves for different volume fractions (graphs in left column \textit{a}, \textit{c}, \textit{e}, \textit{g} and \textit{i}). Hollow squares represent the "`up"' curve from low to high shear rates, filled squares represent the "`down"'curve from high to low shear rates. The solid lines are the results of the schematic model, the dashed line represent the pseudo power law behavior (see Table 2).\break
 Right column: Reduced frequency dependent moduli for different volume fractions (graphs \textit{b}, \textit{d}, \textit{f}, \textit{h} and \textit{j} in right column). Full symbols/solid lines represent \textit{G'}, hollow symbols/dashed lines represent \textit{G''}. Squares mark the data points obtained by the Physica MCR 301, circles with the PAV, triangles up with the TR, the thick lines are the results of the schematic model, the thin lines the results of the microscopic model.\break
Graphs in one row represent the continuous and dynamic measurements at one volume fraction. \textit{a} and \textit{b} were measured at 8.90$\,wt\%$ and 25$\,^{\circ}$C $\phi_{eff}$  = 0.530 , 
\textit{c} and \textit{d} were measured at 7.87$\,wt\%$  and 18$\,^{\circ}$C (circles)  $\phi_{eff}$  = 0.595 , 
\textit{e} and \textit{f} at 8.90$\,wt\%$ and 20$\,^{\circ}$C $\phi_{eff}$  = 0.616 ,
\textit{g} and \textit{h} at 8.90$\,wt\%$  and 19$\,^{\circ}$C $\phi_{eff}$ = 0.625  ,
and \textit{i} and \textit{j} at 8.90$\,wt\%$  and 17$\,^{\circ}$C $\phi_{eff}$ = 0.627 
.\label{Fig:5}  }
    \end{figure*}

    \begin{figure*}
    \addtocounter{figure}{-1}
    \centering
    \includegraphics[width=0.95\textwidth]{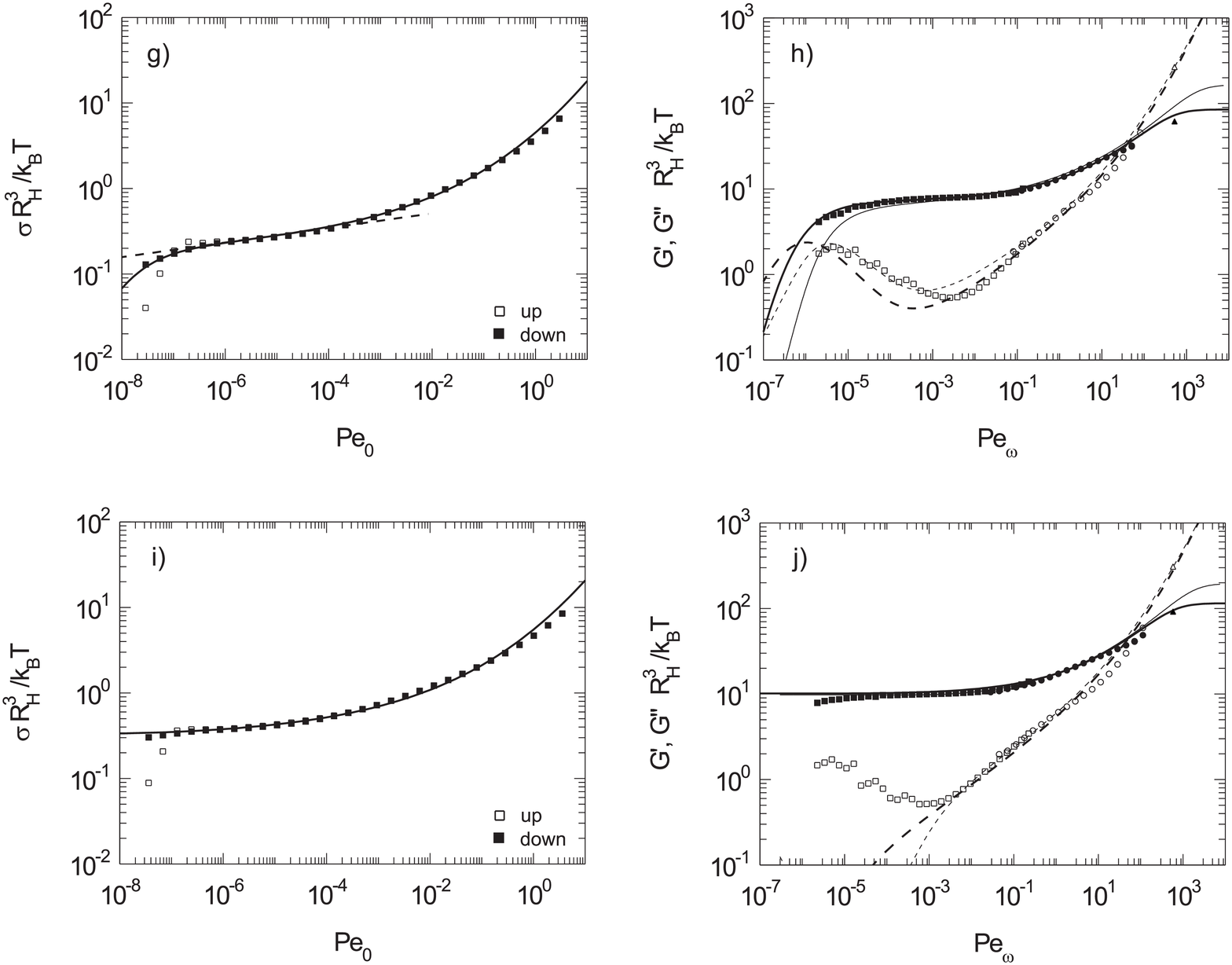}
\caption{Continued.
    \label{Fig:5b}  }
    \end{figure*}

In the following we shall compare the experimental flow curves and
$G'$ and $G''$ with theory for a given effective volume fraction
$\phi_{eff}$. Fig. \ref{Fig:5} gives this comparison for five
different effective volume fractions. On the left-hand side the flow
curves $\sigma(\dot{\gamma})$ are presented as function of the bare
Peclet number $Pe_0$, on the right-hand side $G'$ and $G''$ are
displayed as the function of Pe$_{\omega}$, now calculated with the
respective frequencies $\omega$. Table \ref{table:2} gathers the
respective volume fractions together with the other experimental
conditions. Note that $G'$ and $G''$ have been obtained over nearly
7 orders of magnitude in frequency by combining a Couette rheometer
(single gap, double gap) with the PAV and the torsional resonator.

The generalized shear modulus $g(t, \dot{\gamma})$ (cf. Eq.
\ref{d4}) presents the central theoretical quantity used in this
fit. Within the schematic model, the vertex prefactor $v_{\sigma}$
is kept as a shear-independent quantity. It can easily be obtained
from the stress and modulus magnitudes. Hydrodynamic interactions
enter through $\eta_{\infty}$, which can be obtained from
measurements done at high frequencies, and through $\Gamma$, which
can be obtained via Eq. \ref{d8} from measurements done at high
shear rates. Given these three parameters, both the shapes of the
flow curves as well as the shapes of the moduli $G'$ and $G''$ may
be obtained as  function of the two parameters $\epsilon$ and
$\dot\gamma/\gamma_c$. The former sets the separation to the glass
transition and thus (especially) the longest relaxation time, while
the latter tunes the effect of the shear flow on the flow curve. The
characteristic changes of the shapes of the steady stress and the
moduli are
  dominated by $\epsilon$ and $\dot\gamma/\gamma_c$.

\begin{table}
    \centering
\begin{tabular}{ccccccccccc}
\hline \hline c  &     T  & $\phi_{eff}$ &    $\epsilon$ &
$v_\sigma$ &      $\Gamma$ &    $\gamma_c$ & $\eta_{\infty}$ =
$\eta_{\infty} ^{\omega}$ & $\eta_{\infty}^{\dot{\gamma}}$
\tablenote{calculated by Eq. \ref{d8}}&    $\sigma_0$
\tablenote{extrapolated from the fits using the
$F_{12}^{(\dot{\gamma})}$-model}  &     $\eta_0$ \tablenote{directly
measured for the three lower volume fractions, extrapolated for the
two highest volume fraction from the fits using the $F_{12}^{(\dot{\gamma})}$-model
 } \\

 $\left[wt\%\right]$ & $\left[^{o}C\right]$&
  & &$\left[\frac{k_BT}{R_H^3}\right]$ &$\left[\frac{D_0}{R_H^2}\right]$&&$\left[\frac{k_BT}{D_0R_H}\right]$ &$\left[\frac{k_BT}{D_0R_H}\right]$ &$\left[\frac{k_BT}{R_H^3}\right]$&$\left[\frac{k_BT}{D_0R_H}\right]$ \\

\hline
    8.90 &         25 &
      0.530  & -0.072000  &         18 &         20 &    0.0845  &    0.2250  &    0.6750  &     -       & 1.1345$\cdot 10^{1}$\\

     7.87 &         18 &
      0.595  & -0.003500  &         48 &         50 &    0.1195  &    0.2400  &    0.7200  &     -       & 7.7439$\cdot 10^{3}$ \\

    8.90 &         20 &
      0.616  & -0.000420  &         70 &         80 &    0.1414  &    0.3938  &    0.8313  &      -      & 1.0177$\cdot 10^{6}$ \\

     8.90 &         19 &
      0.625  & -0.000170  &         85 &         90 &    0.1491  &    0.4250  &    0.8972  &        -    & 9.1562$\cdot 10^{6}$ \\

      8.90 &         17 &
     0.627  &  0.000021  &        115 &        120 &    0.1622  &    0.4313  &    0.9104  & 3.3193$\cdot 10^{-1}$ &        -    \\
\hline
\hline
\end{tabular}
\caption{Experimental parameters (concentration $c$, temperature $T$
and calculated $\phi_{eff}$) of the measurements shown in Fig.
\ref{Fig:5}. The fit parameter $v_\sigma$, $\Gamma$, $\gamma_c$ and
$\eta_{\infty}$ of the $F_{12}^{\dot{\gamma}}$ model as well as the
calculated values $\eta_{\infty}^{\dot{\gamma}}$, $\sigma_0$ and
$\eta_0$ are written in terms of the respective normalization.\label{table:2}}
\end{table}

The fitting of the data was done as follows: All measured
quantities, namely $\sigma$, $G'$, and $G''$ were converted to the
respective dimensionless quantities by multiplication with
$R_H^3/k_BT$ where $R_H$ is the hydrodynamic radius at the
respective temperature. Previous work has demonstrated that the
effective volume fraction deriving from $R_H$ is the only decisive
thermodynamic parameter [Crassous \textit{et al.} (2006a, 2008a)].
As already discussed above, $\dot{\gamma}$ and $\omega$ are
converted by $6\pi\eta_s R_H^3/k_BT$ to the respective Peclet
numbers.

Evidently, both the reduced moduli, the Pe number, and the packing
fraction depend on the effective particle volume $R_H^3$. Therefore,
this quantity must be measured with utmost precision. In our
polydisperse sample, actually a distribution of values $R_H^3$
exists, whose variance we determined by disc centrifugation at low
concentration. It is important to note, however, that the size
distribution does not change with temperature. Moreover, in our
experiments close to the glass transition, we may assume the size
distribution to be (almost) density independent.  Therefore, we have
$R_H^3(T)$ as single experimental control parameter, whose small
change upon varying temperature $T$ drives the system through the
glass transition (see also the discussion of Fig. 6).

Subsequently, the fit procedure is started at the smallest volume
fraction and a trial value for the separation parameter $\epsilon$
is chosen. This fit can be done very accurately since the resulting
reduced moduli depend very sensitively on $\epsilon$. Once the
overall shape of both the reduced stress $\sigma R_H^3/k_BT$ and of
the reduced storage and loss modulus has been reproduced by a proper
choice of $\epsilon$, the vertex parameter $v_{\sigma}$ is obtained.
For lower volume fraction the crossing point of $G'$ and $G''$ may
be used, near the glassy state the plateau value of $\sigma$ and
$G'$ can be used. Note that in all steps both sets of data, that is,
the flow curves as well as $G'$ and $G''$ must be fitted with the
same $v_{\sigma}$. Next $\Gamma$ is determined from the dynamic data
which are fully described with this choice of parameters. Given
$\gamma$, the parameter $\gamma_c$ (see Eq. \ref{d2}) is chosen so
that the experimental flow curves are matched by theory. Then
$\eta_0$, $\sigma_0$ and $\eta_\infty^{\dot{\gamma}}$ can be
calculated. The results of these fits are gathered in Table 1. We
reiterate that the final set of parameters must lead to a
description of both the flow curves as well as of the dynamic data.
This requirement stresses the  connection of the equilibrium
structural relaxation at the glass transition to the non-linear
rheology under strong flow, which is one of the  central points of
the ITT-MCT approach.

Fig. \ref{Fig:5} demonstrates that the rheological behavior of a
non-crystallizing colloidal can be modeled in a highly satisfactory
manner by five parameters that display only a weak dependence on the
effective volume fraction of the particles. Increasing the effective
packing fraction drives the system toward the glass transition,
viz.~$\epsilon$ increases with $\phi_{eff}$. Stress magnitudes
(measured by $v_\sigma$) also increase with $\phi_{eff}$, as do high
frequency and high shear viscosities; their difference determines
$\Gamma$. The strain scale $\gamma_c$ remains around the reasonable
value 10$\,\%$. In spite of the smooth and small changes of the
model parameters, the $F_{12}^{\dot{\gamma}}$-model achieves to
capture the qualitative change of the linear and non-linear
rheology. The measured Newtonian viscosity increases by a factor
around $10^5$.  The elastic modulus $G'$ at low frequencies is
utterly negligible at low densities, while it takes a rather
constant value around $10 k_BT/ R^3_H$ at high densities. An
analogous observation holds for the steady state shear stress
$\sigma(\dot\gamma)$, which at high densities takes values around
$0.3 k_BT/ R^3_H$ when measured at lowest shear rates. For lower
densities, shear rates larger by a factor around $10^7$ would be
required to obtain such high stress values.

At volume fractions around 0.5 the suspension is Newtonian at small $Pe_0$ (see Fig. \ref{Fig:5}). Evidently, these flow curves are fully reversible since the suspension is still in the fluid regime. The schematic model provides a perfect fit. Moreover, theory demonstrates in full agreement with experiment that the high-shear viscosities  measured in shear flow and the high-frequency viscosities measured in the linear viscoelastic regime do not agree. As is evident from Eq. \ref{d8}, both $\eta_{\infty}^{\dot{\gamma}}$ and $\eta_{\infty}^{\omega}$ are related by Eq. \ref{d8} and the Cox-Merz rule does not hold for suspensions as expected. Approaching the glass transition leads to a characteristic S-shape of the flow curves and the Newtonian region becomes more and more restricted to the region of smallest $Pe_0$. Concomitantly, a pronounced minimum in $G''$ starts to develop, separating the slow structural relaxation process from faster, rather density independent motions, while $G'$ exhibits a more and more pronounced plateau. At the highest densities (Fig. \ref{Fig:5} panels i and j), the theory would conclude that a yielding glass is formed, which exhibits a finite elastic shear modulus (elastic constant) $G_\infty=G'(\omega\to0)$, and a finite dynamic yield stress, $\sigma^+=\sigma(\dot\gamma\to0)$. The experiment shows, however, that small deviations from this glass like response exist at very small frequencies and strain rates. Description of this ultra-slow process requires extensions of the present ITT-MCT   [Crassous \textit{et al.} (2008a)], where additional physical mechanisms for stress relaxation need to be considered that are neglected here.

\begin{figure*}[t]
    \centering
    \includegraphics[width=0.5\textwidth]{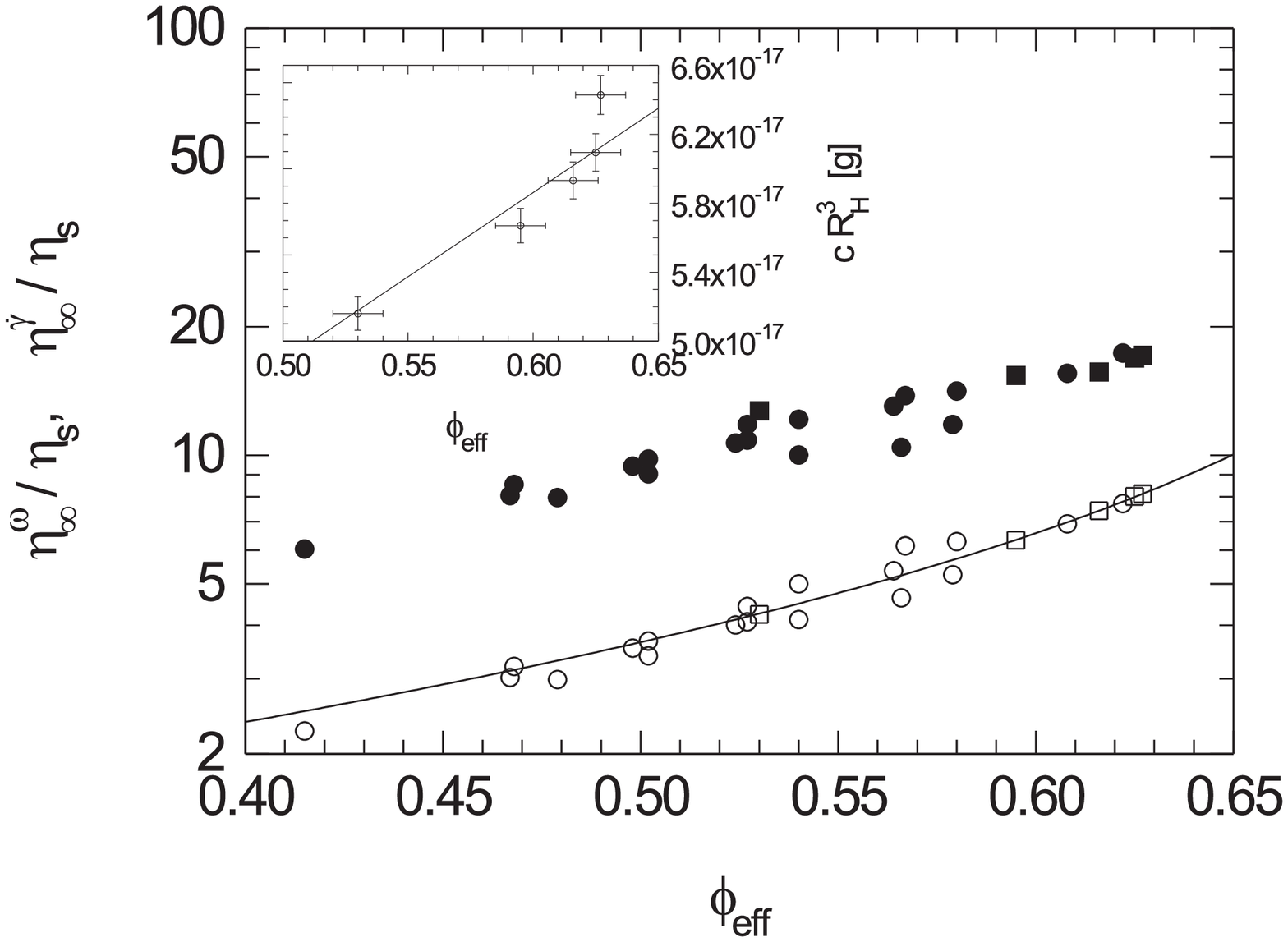}
\caption{Relative viscosities in dependence of the effective volume fraction for a monodisperse latex [Crassous \textit{et al.} (2008a)] (circles) and the polydisperse latex (this work; squares). The hollow symbols show the high frequency viscosities, the filled symbols the high shear viscosities. The relative high frequency viscosity $\eta_{\infty}^{\omega}$ can be described by Lionberger-Russel - equation (see Eq. 4.22 in Lionberger and Russel (1994)) with a reduced effective volume fraction of 11.95$\,\%$. Assuming the same relation for the polydisperse latex under consideration here, $\eta_{\infty}^{\omega}$ was used to determine the effective volume fractions. The $\phi_{eff}$ obtained by this way are plotted in the inset versus the concentration $c$ [$g\cdot mL^{-1}$] times the cubed hydrodynamic radius $R_H^3$ [$m^3$]. The solid line represents the linear scaling of effective volume fraction and $c \cdot R_H^3$ with a slope of 9.77$ \cdot$ 10$^{-17}\,g$ and a intercept of zero. As a further check, the high-shear viscosities are plotted as well as a function of $\phi_{eff}$. The filled circles show this quantity for the narrowly distributed latex [Crassous \textit{et al.} (2008a)] whereas the filled squares denote the repsective quantities for the present system. Good agreement is seen thus corroborating the present way of determining $\phi_{eff}$. \label{Fig:6}  }
    \end{figure*}

\subsection{Power laws}
Often power-law shear thinning is discussed, when the viscosity
is plotted as function of shear rate instead of the flow curves,
viz. stress versus shear rate. For fluid states, the double
logarithmic presentation in Fig. \ref{Fig:5} shows characteristically S-shaped
flow curves, which have a point of inflection. Clearly, this flow curve being a more
sensitive presentation as
the usual $\eta(\dot\gamma)$-plots rules out the existence of a true power
law behavior, which should show up as a straight line in the double
logarithmic plot. Yet, at the point of inflection of the S-curve,
and closely around it, a pseudo-power law may exist. Here
$\eta(\dot\gamma)\sim\dot\gamma^{(p-1)}$, where $p$ is the slope of
the tangent in the flow curve at the inflection point,
$p=d\ln{\sigma}/d\ln{\dot\gamma}$. These tangents are indicated by dashed straight lines
in the panels of Fig. 5. The pseudo-power law that corresponds to the one in the viscosity
can describe the flow curve over more than two decades in shear
rate. The exponents $p$ (see Table \ref{table:3}) change with density, and can be computed
within the schematic model [Hajnal and Fuchs (2008)]. Nevertheless, Fig. \ref{Fig:5} shows clearly that
the description by power laws is by far inferior to the one provided by MCT.

\subsection{Effective volume fraction}
As already lined out in the section Experimental, the higher polydispersity of the suspension used here suppresses crystallization. On the other hand, the precise determination of the volume fractions at which crystallization occurs leads to a precise determination of the effective volume fractions $\phi_{eff}$ [Cheng \textit{et al.} (2002)]. This fact has been used in our previous study using a narrowly distributed thermosensitive system [Crassous \textit{et al.} (2006a, 2008a)]. In this way the zero-shear viscosity as well as the high-frequency and the high-shear viscosities could be obtained as the function of $\phi_{eff}$. The effective volume fraction $\phi_{eff}$ may in principle be determined from the mass balance as lined out in previous studies [Crassous \textit{et al.} (2006a, 2008a)]. However, the error in this determination turned out to be of appreciable magnitude. Hence, we followed the suggestion of Cheng \textit{et al.} (2002) and compared the high-frequency viscosities of the narrowly distributed system studied previously [Crassous\textit{ et al.} (2008a)] with the respective values determined directly for the present system. Fig. \ref{Fig:6} displays the resulting dependence of the high-frequency viscosities (hollow symbols) and the high-shear viscosities (filled symbols) on $\phi_{eff}$. The relative high frequency viscosity $\eta_{\infty}^{\omega}$ can be described by equation 4.22 in [Lionberger and Russel (1994)] with a reduced effective volume fraction of 11.95$\,\%$ as opposed to the actual value of $\phi_{eff}$, the volume transition neglegting a partial draining. The marked reduction of the volume fraction at high frequencies has already been detected in a previous study of a similar system [Deike \textit{et al.} (2001)] and was explained by a partial draining of the thermosensitive network. The resulting correlation between the effective volume fractions and $\eta_{\infty}^{\omega}$ of the narrowly distributed systems can then be used to determine $\phi_{eff}$ of the present system. The values thus obtained are gathered in Table 1. Moreover, the inset of Fig. \ref{Fig:6} demonstrates that there is a linear relationship between these $\phi_{eff}$ and $c R_H^3$ as discussed above. A further check of the data is provided by a comparison of the high-shear viscosities $\eta_{\infty}^{\dot{\gamma}}$ for both systems. Fig. \ref{Fig:6} demonstrates that both sets of data agree within the given limits of error. This fact demonstrates that the present procedure leads to reliable values of $\phi_{eff}$.
\\

\subsection{Microscopic MCT calculations}
While schematic models  provide a handy description of the universal
features close to the glass transition and the shear melting of
glasses, the microscopic formulation of ITT-MCT actually aims to
predict from first principles the nonlinear rheology starting with
the underlying particle interactions; see the Appendix for a short
summary. Because of the anisotropy of the structure under shear, it
has not been possible yet to solve the ITT-MCT equations without
approximations. Only solutions of schematic or isotropically
averaged models are available. Yet, without shear the equations of
quiescent MCT [G\"otze 1991] can be solved readily and can be
compared to the frequency dependent moduli within linear response
regime. This has been done considering the monodisperse system in
Ref. [Crassous \textit{et al.} (2008a)], and after determination of the
effective packing fraction we could perform identical calculations
for the polydisperse system; see Ref. [Crassous \textit{et al.}
(2008a)] for details of the calculation which use the Percus-Yevick
approximation to the structure factor of hard spheres. Here, of
course the simplification to do monodisperse MCT calculations to
compare with the measurements in the polydisperse system adds to the
errors in the comparisons. The adjustable parameters, mainly the
effective collective short time diffusion coefficient and the
effective packing fraction setting the separation to the glass
transition are reported in Table \ref{table:3}. Quite good agreement with the
data can be achieved, with parameters that vary regularly with
density and take quite reasonable values. Interestingly, the
polydisperse system has larger moduli by a factor around  two than
the monodisperse system at the same effective packing fraction.

\begin{table}[h]
\centering
\begin{tabular}{cccccccc}
\hline
\hline
         $c$ &          $T$ & $\phi_{eff}$ & $\epsilon$   &$\phi$ &   $c_y$ & $D_s/D_0$ & $p$ \\

    $[wt\%]$ & $[^{o}C]$ &            &  & &          &  &          \\
\hline
     8.90  &     25  &     0.530  &      -0.10000 & 0.4643 &    2.3  &      0.3  & 0.631\\

     7.87  &     18  &     0.595  &      -0.00794 & 0.5118  &   2.3  &      0.3  & 0.248\\

     8.90  &     20  &     0.616  &      -0.00126 & 0.5153 &    2.3  &      0.3  &0.117\\

     8.90  &     19  &     0.625  &      -0.00100 & 0.5154  &   3.0  &      0.3  & 0.0852\\

     8.90  &     17  &     0.627  &      0.00158 & 0.5167   &  3.5  &      0.3  &\\
\hline
\hline
\end{tabular}
\caption{Experimental parameters (concentration $c$, temperature $T$
and experimental effective volume fraction $\phi_{eff}$ with
$\phi_g$ = 0.58. The parameters of the microscopic calculations
(separation parameter $\epsilon$, volume fraction $\phi$ with
$\phi_g$ = 0.5159, $c_y$ and $D_S/D_0$, where $D_0=k_BT/(6\pi\eta_s
R_H)$), as shown in Fig. \ref{Fig:5} are listed here. Furthermore, the power law exponents $p$ from the dashed lines in the flowcurves of Fig. \ref{Fig:5} are given in this table.
\label{table:3}}
\end{table}

\section{Conclusion}
The fluid-to-glass transition in colloidal suspensions leads to
characteristic changes of the flow curves and concomitantly of $G'$
and $G''$ as the function of frequency. Here we demonstrated that
both sets of data can be quantitatively modeled by a single
generalized modulus $g(t,\dot\gamma)$. Moreover, $g(t,\dot\gamma)$
can be fully described by mode-coupling theory (MCT). Hydrodynamic
interaction enters only through the high-shear and the
high-frequency viscosities $\eta_{\infty}^{\dot{\gamma}}$ and
$\eta_{\infty}^{\omega}$, respectively. MCT demonstrates that the
latter quantities are not identical but closely related to each
other (see the discussion of Eq. \ref{d8}).  Small deviations
between data and theory are only seen for $G''$ in the immediate
vicinity of the glass volume fraction and only at the lowest Peclet
numbers $Pe_{\omega}$. This finding can be assigned to a very slow
relaxation that is not captured by MCT.


\begin{theacknowledgments}
  We thank D. Hajnal and O. Henrich for valuable help, and the Deutsche Forschungsgemeinschaft, Forschergruppe 608 "Nichlineare Dynamik", Bayreuth, and IRTG 667 "Soft Condensed Matter Physics", Konstanz, for financial support.
\end{theacknowledgments}

\bigskip

\appendix{\bf Appendix }
\bigskip

The ITT approach, which generalizes the MCT of the glass transition
to colloidal dispersions under stationary shear [Fuchs and Cates
(2002); ibid.~(2008)], consists of equations of motion for a density
correlator $\Phi_{\bf q}(t)$ which describes structural
rearrangements, and approximated Green-Kubo formulae relating stress
relaxation to decay of density fluctuations. The approach was
described in [Crassous \textit{et al.} (2008a)], and the used equations can be
summarized as follows.

The density correlator obeys an equation of motion containing a
retarded friction kernel which arises from the competition of
particle caging and shear advection of fluctuations
\begin{equation}
\partial_t \Phi_{\bf q}(t) + \Gamma_{\bf q}(t) \; \left\{
\Phi_{\bf q}(t) + \int_0^t dt'\;  m_{{\bf q}(t')}(t-t') \;
\partial_{t'}\,  \Phi_{\bf q}(t') \right\} = 0 \; . \end{equation}
with the advected wavector ${\bf q}(t)$, whose component along the
gradient direction varies with time, $q_y(t)=q_y(0) - \gd t\; q_x$
because of the affine component of the particle motion; while $q_x$
and $q_z$ remain time-independent. The density correlator is
normalized $\Phi_{\bf q}(t=0)=1$ . Its initial decay is described by
$\Gamma_\qb(t)=  D_s q^2(t)/ S_{q(t)}$, with $D_s$ the collective
short time diffusion coefficient, which may depend on hydrodynamic
interactions, and thus differ from the Stokes-Einstein-Sutherland
value. The equilibrium structure factor $S_q$, encodes the particle
interactions and introduces the experimental control parameters like
density and temperature. The generalized friction kernel $m_{{\bf
q}'}(t)$, which is an autocorrelation function of fluctuating
stresses, is approximated from the structural rearrangements
captured in the density correlators
\begin{equation}  m_{\bf q'}(t) = \frac{1}{2N}
\sum_{\bf k} \frac{S_{q'(t)}\, S_{k}\, S_{p}}{{q'}^2(t)\;
{q'}^{2}}\;
 V_{\qb'\kb\pb}(t)\, V_{\qb'\kb\pb}(0)\;
 \Phi_{\bf k}(t)\;  \Phi_{\bf p}(t)\; ,
\end{equation} where the  vertices are given by the equilibrium structure factor
\begin{equation}
V_{\bf qkp}(t) =  {\bf q}(t)\cdot \left( {\bf k}(t) \, n c_{k(t)} +
{\bf p}(t)\, n c_{p(t)} \right)\; \delta_{\qb,\kb+\pb} \;
,\end{equation} with $n$ the particle density, and $c_k$ the
Ornstein-Zernicke direct correlation function $c_k=(1-1/S_k)/n$.
Similarly, the generalized shear modulus is approximated assuming
that stress relaxations can be computed from integrating the
transient density correlations
\begin{equation}
g(t,\dot\gamma) = \frac{k_BT}{2}
\int\!\!\frac{d^3k}{(2\pi)^3}\;
\frac{k_x^2k_yk_y(-t)}{k\, k(-t)}\;
\frac{S'_kS'_{k(-t)}}{S^2_{k}}\; \Phi^2_{\kb(-t)}(t)\; ,
\end{equation}
where $S'_k=\partial S_k/\partial k$.

The schematic $F_{12}^{(\dot\gamma)}$-model described above follows
from a simplification of these equations. It mimicks  using $v_1$
and $v_2$ the increase of the vertices $V_{\qb'\kb\pb}$ with
sharpening structural correlations in $S_q$ , which captures the
cage effect causing a glass transition [G\"otze (1991), G\"otze and
Sj\"ogren (1992)]. It mimicks with $1/(1+(\dot\gamma t/\gamma_c)^2)$
the shear-driven decorrelation, which can be recognized from the
vanishing of the memory kernel at long times under shear: $m_{{\bf
q'}}(t \dot\gamma \gg 1) = 0$ and $m_{{\bf q'}}(t)-
m_q(t)^{(\dot\gamma=0)}={\cal O}((\dot\gamma t)^2)$ in the
microscopic ITT equations [Fuchs and Cates (2002)]. Without shear,
$\dot\gamma\equiv0$, Eqs. (8-11) become  the familiar MCT equations
which can readily be solved after discretization, as has been done
in previous studies [Crassous \textit{et al.} (2008a)]. The thin dashed lines
in Fig. 5 labeled microscopic model result from these calculations,
with parameters listed in Table 2. But under shear the equations
become anisotropic and require more severe approximations, like for
example using the simple schematic $F_{12}^{(\dot\gamma)}$-model.

In detail, the equations were solved for a dispersion of
monodisperse hard spheres with structure factor $S_k$ taken from the
Percus-Yevick approximation. The wavevector integrals in the
microscopic model Eqs.~(8-12) were discretized according to Ref.
[Crassous et al. (2008)] using $M=600$ wavevectors chosen from
$k_{\rm min}=0.05/R_H$ up to $k_{\rm max}=59.95/R_H$ with separation
$\Delta k = 0.1/R_H $. Time was discretized with initial step-width
$dt=2\,10^{-7} R_H^2/D_s$, which was doubled each time after 400
steps. Because the first principles calculations carry quantitative
errors, which would mask the qualitative comparison, two adjustments
of parameters were performed. The magnitude of the linear modulus is
too small in theory by a factor $c_y$ which is included in Table 2.
The computed position of the glass transition is also somewhat to
small, $\phi_g=0.5159$ in MCT relative to $\phi_g=0.58$ in
experiment. Therefore the relative separation,
$\varepsilon=(\phi-\phi_g)/\phi_g$ was used as adjustable parameter
quantifying the distance to the glass transition; the fitted values
are also contained in Table 2.



\bibliographystyle{aipproc}   

\section*{References}

\small{
\begin{description}

\item[]Banchio, A. J., J. Bergenholtz, G. N{\"a}gele, ``Viscoelasticity and generalized Stokes-Einstein relations of colloidal dispersions.'', J. Chem. Phys. \textbf{111}, 8721 (1999).


\item[]Callaghan, P. ``Rheo NMR and shear banding.'' Rheol. Acta, \textbf{47}, 243 (2008).

\item[]Cheng, Z., J. Zhu, P. M. Chaikin, S.-E. Phan, W. B. Russel, "`Nature of the divergence in low shear viscosity of colloidal hard-sphere dispersions"', Phys. Rev. E \textbf{65}, 041405 (2002).

\item[]Crassous, J.~J., R. Regisser, M. Ballauff and W. Willenbacher, "`Characterization of the viscoelastic behavior of complex fluids using the piezoelastic axial vibrator"', J. Rheol. \textbf{49}, 851 (2005).

\item[]Crassous, J.~J., M. Siebenb{\"u}rger, M. Ballauff, M. Drechsler, O. Henrich, M. Fuchs, ``Thermosensitive core-shell particles as model systems for studying the flow behavior of concentrated colloidal dispersions'', J. Chem. Phys. \textbf{125}, 204906 (2006a).

\item[]Crassous, J. J., M. Ballauff, M. Drechsler, J. Schmidt, Y. Talmon, ``Imaging the Volume Transition in thermosensitive Core-Shell Particles by Cryo-Transmission Electron Microscopy'' Langmuir \textbf{22}, 2403 (2006b).

\item[]Crassous, J.~J., M. Siebenb{\"u}rger, M. Ballauff, M. Drechsler, D. Hajnal, O. Henrich, M. Fuchs, ``Shear stresses of colloidal dispersions at the glass transition in equilibrium and in flow'', J. Chem. Phys. \textbf{128}, 204902 (2008a).

\item[]Crassous, J. J., A. Wittemann, M. Siebenbrger, M. Schrinner, M. Drechsler, M. Ballauff, ``Direct Imaging of Temperature-Sensitive Core-Shell Latexes by Cryogenic Transmission Electron Microscopy'', Colloid Polym. Sci. \textbf{286}, 805 (2008b).

\item[]Deike, I., M. Ballauff, N. Willenbacher, A. Weiss, "`Rheology of thermosensitive latex particles including the high-frequency limit"', J. Rheol. \textbf{45}, 709 (2001).

\item[]Dhont J, Briels W ``Gradient and vorticity banding.'' Rheol. Acta \textbf{47}, 257 (2008).

\item[]Dingenouts, N., Ch. Norhausen and M. Ballauff,  "`Observation of the Volume Transition in Thermosensitive Core-Shell Latex Particles by Small-Angle X-ray Scattering"', Macromolecules, \textbf{31}, 8912 (1998).

\item[]Fritz, G., W. Pechhold, N. Willenbacher and J.~W. Norman, "`Characterizing complex fluids with high frequency rheology using torsional resonators at multiple frequencies"', J. Rheol. \textbf{47}, 303 (2003).

\item[]Fuchs, M. and M.~E. Cates, "`Theory of Nonlinear Rheology and Yielding of Dense Colloidal Suspensions"', Phys. Rev. Lett. \textbf{89}, 248304 (2002).

\item[]Fuchs, M. and M.~E. Cates, "`Schematic models for dynamic yielding of sheared colloidal glasses"', Faraday Disc. \textbf{123}, 267 (2003).

\item[]Fuchs, M. and M. Ballauff, ``Flow curves of dense colloidal suspensions: Schematic model analysis of the shear-dependent viscosity near the colloidal glass transition'', J. Chem. Phys. \textbf{122}, 094707 (2005).

\item[]Fuchs, M. and M.~E. Cates, "Mode coupling approximations for Brownian particles in homogeneous steady shear flow", in preparation (2008)

\item[]G{\"o}tze, W. "` Some aspects of phase transitions described by the self consistent current relaxation theory"', Z. Phys. B \textbf{56}, 139 (1984).

\item[]G{\"o}tze, W. in \textit{Liquids, Freezing and Glass Transition}, edited by J. P. Hansen, D. Levesque, and J. Zinn-Justin, Session LI (1989) of Les Houches Summer Schools of Theoretical Physics, (North-Holland, Amsterdam, 1991), 287.

\item[]G{\"o}tze, W. and L. Sj{\"o}gren, "`Relaxation processes in supercooled liquids"', Rep. Prog. Phys. \textbf{55}, 241 (1992).

\item[]Hajnal, D. and M. Fuchs, "`Flow curves of colloidal dispersions close to the glass transition asymptotic scaling laws in a schematic model of mode coupling theory."', Cond. Mat.  1-14, arXiv:0807.1288v1 [cond-mat.soft] (2008).

\item[]Helgeson, M.~E., N.~J. Wagner, D. Vlassopoulos, ``Viscoelasticity and shear melting of colloidal star polymer glasses'', J. Rheol. \textbf{51}, 297 (2007).

\item[]Heymann, L. and N. Aksel, ``Transition pathways between solid and liquid state in suspensions.'', Phys. Rev. E \textbf{75}, 021505 (2007).

\item[]Isa, L., R. Besseling, W.~C.~K. Poon, "`Shear zones and wall slip in the capillary flow of concentrated colloidal suspensions"', Phys. Rev. Lett. \textbf{98}, 198305 (2007).

\item[]Larson, R.~G. \textit{The Structure and Rheology of Complex Fluids} (Oxford University Press, New York, 1999).

\item[]Lionberger, R. A. and W. B. Russel, "`High frequency modulus of hard sphere colloids"', J. Rheol. \textbf{38}, 18851908 (1994).

\item[]Manneville, S. "`Recent experimental probes of shear banding."' Rheol. Acta \textbf{47}, 301 (2008).

\item[]Meeker, S. P., W. C. K. Poon, P. N. Pusey, ``Concentration dependece of the low shear viscosity of suspensions of hard-sphere colloids'', Phys. Rev. E \textbf{55}, 5718 (1997).

\item[]Olmsted, P. D. ``Perspectives on shear banding in complex fluids.'' Rheol. Acta \textbf{47}, 283 (2008).

\item[]Purnomo, E.~H., D. van den Ende, J. Mellema and F. Mugele, "`Linear viscoelastic properties of aging suspensions"', Europhys. Lett. \textbf{76}, 74 (2006).

\item[]Purnomo, E.~H., D. van den Ende, J. Mellema and F. Mugele, ``Rheological properties of aging thermosensitive suspensions'', Phys. Rev. E \textbf{76}, 021404 (2007).

\item[]Russel, W.~B., D.~A. Saville, and W.~R. Schowalter, \textit{Colloidal Dispersions} (Cambridge University Press, New York, 1989).

\item[]Senff, H., W. Richtering, Ch. Norhausen, A. Weiss, M. Ballauff, "`Rheology of a Temperature Sensitive Core-Shell Latex"', Langmuir \textbf{15}, 102 (1999).

\item[]Seth, Y. R., M. Cloitre, R. T. Bonnecaze, ``Elastic properties of soft particle pastes'', J. Rheol. \textbf{50}, 353 (2006).

\item[]van Megen, W. and P.~N. Pusey, "`Dynamik light scattering study of the glass transition in a colloidal suspension"', Phys. Rev. A, Phys. Rev. A \textbf{43}, 5429 (1991).

\item[]van Megen, W. and S.~M. Underwood, "`Glass-transition in colloidal hard spheres - measurement and mode-coupling -theory analysis of the coherent intermediate scattering function"', Phys. Rev. E \textbf{49}, 4206 (1994).

\item[]Varnik, F.  and O. Henrich, "Yield Stress Discontinuity in a Simple Glass",  Phys. Rev. B {\bf 73}, 174209 (2006).
\end{description}}

\end{document}